\documentclass[preprint,tightenlines,aps,prd,showpacs,nofootinbib]{revtex4}
\usepackage{graphicx}
\usepackage{amsmath}
\usepackage{amssymb}
\usepackage{bm}
\begin{document}
\title{\mbox{}\\[10pt] Parton Model Calculation \\ 
                       of Inclusive Charm Production  \\
                       by a Low-energy Antiproton Beam}
\author{Pierre Artoisenet}
\affiliation{Center for Particle Physics and Phenomenology,
Universit\'e Catholique de Louvain, B1348 Louvain-la-Neuve, Belgium}
\author{Eric Braaten}
\affiliation{Physics Department, Ohio State University, Columbus, OH
  43210, USA}
\affiliation{Bethe Center for Theoretical Physics,
	Universit\"at Bonn, 53115 Bonn, Germany}
\date{\today}
\begin{abstract}
The cross section for inclusive charm production by a low-energy 
antiproton beam is calculated using  the parton model and 
next-to-leading order perturbative QCD.
For an antiproton beam with a momentum of 15 GeV, 
the charm cross section at next-to-leading order in the QCD coupling constant 
changes by more than an order of magnitude as the charm quark mass is varied
from 1.3 to 1.7~GeV.  The variations can be reduced by demanding that 
the same value of the charm quark mass give the measured 
charm cross sections for fixed-target experiments with a proton beam.
The resulting estimate for the charm cross section 
from a low-energy antiproton beam
is large enough to allow the study of charm meson mixing. 
\end{abstract}



\maketitle

Flavor physics in the charm sector is an aspect of  
elementary particle physics that has not been fully explored.
The Standard Model of particle physics predicts the rates 
for charm meson mixing and for CP violation in charm meson decays 
to be very small, which makes the charm sector a promising window 
into new physics beyond the Standard Model \cite{Artuso:2008vf}.
Mixing of the neutral charm mesons $D^0$ and $\bar D^0$ was 
finally observed by several experiments 
in 2007~\cite{Braaten:2007sh,Staric:2007dt,CDF:2007uc}.
Searches for CP violation in the charm sector would require significantly 
larger data samples.  One possible source of large data samples of
flavor-tagged charm mesons is a low-energy antiproton beam, 
such as the High Energy Storage Ring (HESR) planned at the GSI laboratory.
Quantitative predictions for the charm cross section at such a facility 
are needed in order to assess the prospects for studying charm
flavor physics.

There have been very few previous estimates of the charm production rate
by a low-energy antiproton beam.  Braaten estimated the exclusive 
cross section for $D^{*0} \bar D^0$ in low-energy  $\bar p p$ collisions 
by scaling measured cross sections for $K^{*-} K^+$ \cite{Braaten:2007sh}.
He obtained a surprisingly large cross section of 360~nb 
at a center-of-mass energy of 5.71~GeV.
Titov and K\"ampfer have used a Regge model based on baryon exchange
to predict the differential cross sections and longitudinal asymmetries 
for the exclusive production of pairs of charm mesons \cite{Titov:2008yf},
but they did not calculate cross sections integrated over angles.
In this paper, we use the parton model 
and perturbative QCD to calculate the inclusive charm 
cross section from a low-energy antiproton beam.

The parton model and perturbative QCD  
have been used extensively to calculate cross sections for
charm quarks, bottom quarks, and top quarks in hadron collisions.
In the case of charm production, the applications have ranged from 
fixed-target experiments with proton beams with center-of-mass 
energy $\sqrt{s}$ below 10~GeV to 
Fermilab's Tevatron $\bar p p$ collider and CERN's Large Hadron Collider
with $\sqrt{s} = 2$~TeV and 14~TeV, respectively.
The parton model predictions for top quark production and for the 
transverse momentum distributions for bottom quarks and charm quarks are well 
tested at the Tevatron.  The parton model was used to predict the 
charm cross section in $p p$ collisions at $\sqrt{s} = 200$~GeV
at Brookhaven's Relativistic Heavy Ion Collider (RHIC) \cite{Cacciari:2005rk}.
We will use the parton model to make the much larger extrapolation 
down to $\bar p p$ collisions with center-of-mass energies as low as 
5.5~GeV.

A convenient source of flavor-tagged charm mesons  
in high energy physics experiments is the production of 
the charm mesons $D^{*\pm}$. 
In the parton model, the inclusive cross section for producing 
$D^{*+}$ in $\bar p p$ collisions
can be expressed as
\begin{eqnarray}
\sigma[\bar p(P_A) p(P_B) \to D^{*+}  + X] &=&
\sum_{ij} \int_0^1 d x_1 \, f_{i/\bar p}(x_1) 
\int_0^1 d x_2 \, f_{j/p}(x_2) \, 
\nonumber
\\ && ~~~~~
\times \hat \sigma[ i(x_1 P_A) j (x_2 P_B) \to c + X] \, P_{c \to D^{*+}},
\label{sigma}
\end{eqnarray}
where $P_{c \to D^{*+}}$ is the fragmentation probability for a 
charm quark to hadronize into a $D^{*+}$ meson.
At low energies, the dominant contributions 
to the sums over the types of partons come from 
$(i,j) = (\bar u,u)$ and $(\bar d,d)$.
The parton distributions $f_{i/\bar p}$ and $f_{j/p}$ depend 
on a factorization scale $\mu_f$.  The parton cross section
$\hat \sigma$ can be calculated as a power series expansion in 
$\alpha_s(\mu_r)$, with coefficients that depend on $\mu_f$ 
and on the renormalization scale $\mu_r$. 

The lowest energy experiment in which the fragmentation probability 
$P_{c \to D^{*+}}$ has been measured accurately 
is $e^+ e^-$ annihilation at a center-of-mass energy of 10.6~GeV.
The Belle Collaboration has made precise measurements of the 
inclusive cross sections for $D^0$, $D^+$, $D_s^+$,
$\Lambda_c^+$, $D^{*0}$, and $D^{*+}$ \cite{Seuster:2005tr}.
Using the constraint that the fragmentation probabilities 
for $D^{0}$, $D^{+}$, $D_s^+$, and $\Lambda_c^+$ add up to 1,
we can determine the fragmentation probability for $D^{*+}$ 
at that energy scale to be $P_{c \to D^{*+}} = 24\%$.
We will use this as an estimate of $P_{c \to D^{*+}}$ 
in low-energy $\bar p p$ collisions.

To assess whether the extrapolation of the parton model 
calculation to center-of-mass energies as low as 
5.5~GeV is plausible, 
we first consider charm production in $e^+ e^-$ annihilation. From 
the charm meson threshold at $\sqrt{s}=3.73$~GeV
up to about 4.6~GeV, the cross section varies dramatically
due to the charmonium resonances $\psi(3730)$, $\psi(4040)$, 
$\psi(4160)$, and $\psi(4415)$.  Only above 4.6~GeV is 
the cross section smooth enough to agree reasonably well
with perturbative QCD.  However {\it quark-hadron duality}
can be exploited to extend the region of applicability 
of perturbative QCD to significantly lower energies  
by smearing over the energy \cite{Poggio:1975af}.
The duality holds between the perturbative charm quark cross section 
and the sum of the cross sections for charm hadrons and 
for charmonium states below the charm threshold.
Since the perturbative cross section is smooth without any local minima,
the smeared hadron cross section should also be smooth.
The minimum smearing interval is the spacing between 
radial excitations of the charmonium states, which
is about 400~MeV.  For the perturbative charm quark cross section 
to be a good approximation to the charm hadron cross section alone, 
the energy must be large enough that the smeared cross section
receives no significant contributions from the charmonium states
below the charm threshold and in particular from the $\psi(3686)$.
This requires the energy at the center of the smearing interval
to be greater than about 3.9~GeV.

We now compare inclusive charm production in $\bar p p$ collisions
and $e^+ e^-$ annihilation. One difference is in the charmonium resonances 
that can be produced.  The annihilation of  $e^+ e^-$ can only produce
resonances with quantum numbers $1^{--}$. In contrast, $\bar p p$ collisions
can produce charmonium resonances with any of the possible quantum numbers 
$J^{PC}$.  Potential models predict the spacing between radial excitations 
in each angular momentum channel to be about 400~MeV.  Thus 
smearing in the $\bar p p$ center-of-mass energy by that amount
might be sufficient to get a smooth cross section above 
the $D \bar D$ threshold in each $J^{PC}$ channel.  
The sum over $J^{PC}$ will help to further smooth out the cross section.
The parton model suggests that it may not be necessary to 
smear over the $\bar p p$ center-of-mass energy,
because the integration over the momentum fractions 
$x_1$ and $x_2$ of the colliding 
partons automatically provides some smearing in the energy. 
To get a smooth cross section, it may be sufficient to demand 
that this integration provides smearing in the center-of-mass energy 
of the colliding partons by at least 400~MeV.

There is an important difference between $e^+ e^-$ annihilation
and $\bar p p$ collisions in the parton processes that create 
the $c \bar c$ pair and therefore in the hadronization processes 
that produce the charm mesons.  The annihilation of $e^+ e^-$  
creates a $c \bar c$ pair and gluons whose overall color state is a singlet. 
In $\bar p p$ collisions, the colliding partons have color connections 
to the remnants of the $\bar p$ and $p$.
The $c \bar c$ pair and the gluons that are created by the collision 
therefore also have color connections to the remnants of the $p$ and $\bar p$.
Part of the energy associated with the constituent mass of the 
light quark in the charm meson may be provided by those remnants.
Thus it is possible for a parton collision with center-of-mass energy 
$\sqrt{\hat s}$ below the charm meson threshold to produce
charm mesons.  

Based on both the analogy with $e^+ e^-$ annihilation as well as the differences, 
we propose the following two conditions as plausible criteria 
for the validity of the parton model calculation of charm production:
\begin{description}
\item{{\bf A}.}
The integration over parton momentum fractions should provide
smearing of at least 400~MeV in the invariant mass $\sqrt{\hat s}$
of the colliding partons.
\item{{\bf B}.}
The dominant contributions should come from $\sqrt{\hat s}$
above the $D \bar D$ threshold at 3.73~GeV.  
\end{description}
If the center-of-mass energy is too low for the
criteria {\bf A} and {\bf B} to be satisfied, 
the parton model calculation can still be used as a 
phenomenological model for extrapolating the charm cross section
to low energies.

\begin{figure}[t]
\includegraphics[width=0.8\textwidth,clip=true]{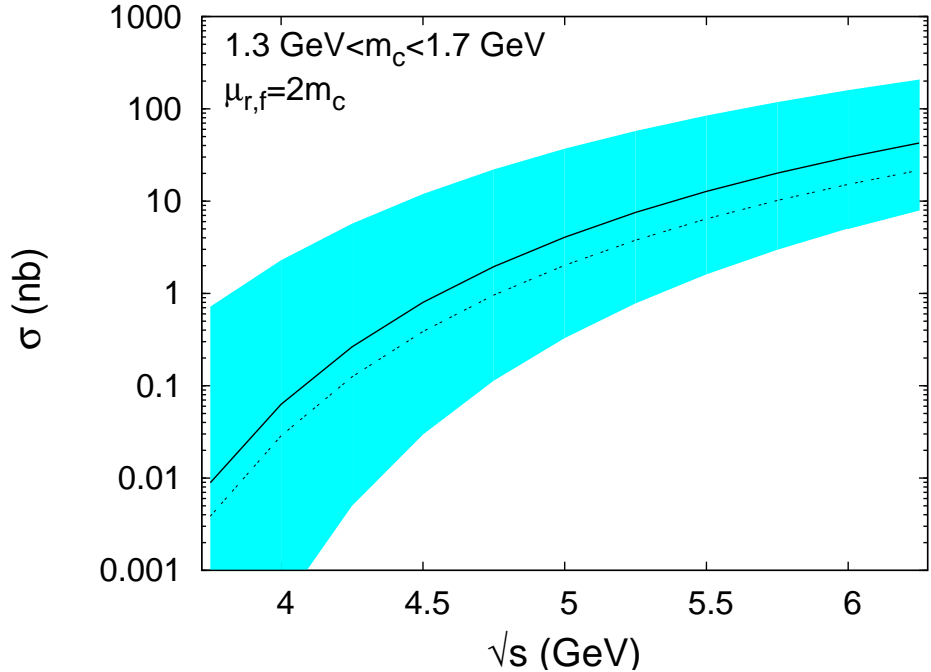}
\caption{Cross section for inclusive charm production in 
$\bar p p$ collisions as a function of the center-of-mass energy
$\sqrt{s}$. The solid line is the NLO result  for $m_c = 1.5$~GeV 
and $\mu_r = \mu_f = 2 m_c$.
The shaded band corresponds to varying $m_c$ from 1.3 and 1.7~GeV.
The dashed line is the LO result.} 
\label{Fig:sig-mc}
\end{figure}

\begin{figure}[t]
\includegraphics[width=0.8\textwidth,clip=true]{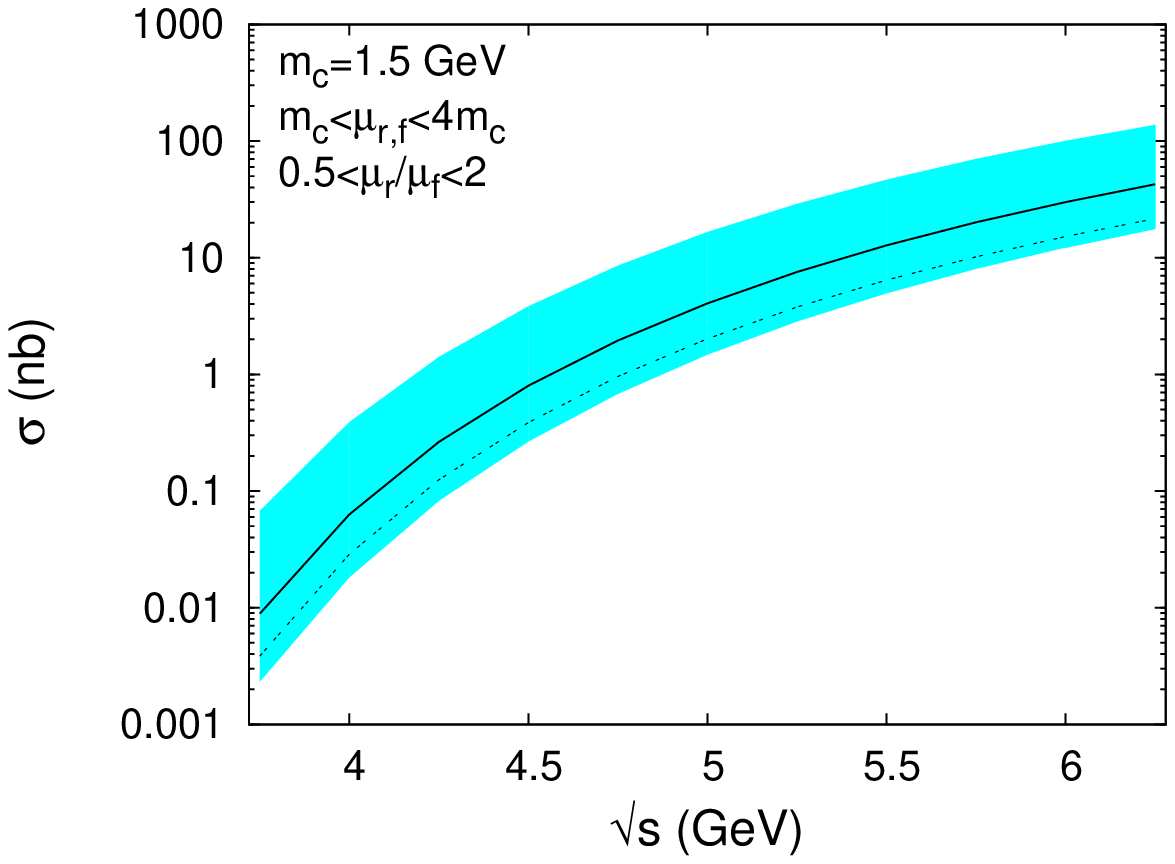}
\caption{Cross section for inclusive charm production in 
$\bar p p$ collisions as a function of the center-of-mass energy
$\sqrt{s}$. The solid line is the NLO result for $m_c = 1.5$~GeV 
and $\mu_r = \mu_f = 2 m_c$.  The shaded band corresponds to varying 
$\mu_r$ and $\mu_f$ from $m_c$ to $4 m_c$ with $\frac12 < \mu_f/\mu_r < 2$.
The dashed line is the LO result.} 
\label{Fig:sig-mu}
\end{figure}

To calculate the charm cross section in Eq.~(\ref{sigma}),
we use the well-known next-to-leading order 
(NLO) calculations of the parton cross sections for heavy-quark 
production~\cite{Nason:1987xz,Beenakker:1988bq,Beenakker:1990maa}.
The parameters in the charm cross section are the charm quark mass
and the renormalization and factorization scales.
We do not carry out any resummation of logarithms of the transverse momentum,
because these logarithms are not large at this low energy.
We also do not carry out any resummation of the threshold logarithms of 
$\hat s - 4 m_c^2$ \cite{Kidonakis:2002vj,Kidonakis:2004qe}, 
where $\hat s$ is the invariant mass of the colliding partons,
although these logarithms can be large.
The inclusive cross section for
$\bar p p \rightarrow c \bar c+X$ is calculated 
using the Monte-Carlo  program 
MCFM~\cite{Campbell:1999ah}. 
The implementation of the cross section at NLO makes
use of the matrix elements calculated in Ref.~\cite{Nason:1987xz}. 
This computation is carried out 
in the $\overline{\textrm{MS}}$ scheme with $n_{\textrm{f}}=3$
light flavors of quarks in the initial state.
At LO and NLO, we use the parton distribution sets CTEQ6L1
and CTEQ6M, respectively~\cite{Pumplin:2002vw}.
These sets have been constructed for $n_{\textrm{f}}=5$ initial flavors, which
leads to a suppression of the gluon density
compared to the case $n_{\textrm{f}}=3$.
%
%
%
At NLO accuracy, the correction term 
that must be added to compensate for the different number of flavors 
is proportional to the gluon-initiated Born-level cross
section and depends logarithmically on 
the ratio $\mu_f/m_c$~\cite{Cacciari:1998it}.
Since our factorization scale is 
comparable to $m_c$, the correction term 
proves to be small and can be disregarded compared to 
other theoretical errors.

The parameters $m_c$, $\mu_r$, and $\mu_f$ and their theoretical 
uncertainties are determined as follows.
We choose their central values to be $m_c = 1.5$ GeV and
$\mu_r = \mu_f = 2 m_c$.  Our central value of $\mu_r = \mu_f$  
is twice that used in Ref.~\cite{Cacciari:2005rk}, 
because this choice decreases 
the sensitivity to the scales at the low energies we consider.
The scale $2 m_c$ is a natural choice, 
because this is the minimal invariant mass of the 
virtual gluon in the dominant leading-order parton process 
$\bar q q \to c \bar c$.
We obtain the error band from the charm quark mass by setting
$\mu_r = \mu_f = 2 m_c$ and allowing $m_c$ to range from $1.3$ to $1.7$ GeV.
We obtain the error band from the scales by setting $m_c = 1.5$~GeV
and allowing $\mu_r$ and $\mu_f$ to range from $m_c$ to $4 m_c$
with $\frac12 < \mu_f/\mu_r < 2$.

Our results for the inclusive charm cross section are shown in 
Figs.~\ref{Fig:sig-mc} and \ref{Fig:sig-mu} as functions of the 
center-of-mass energy $\sqrt{s}$.
Two values of $\sqrt{s}$ that are of particular interest are 4.11~GeV, 
which corresponds to the maximum antiproton momentum of 8~GeV 
in Fermilab's accumulator ring,
and 5.47~GeV, which corresponds to the maximum antiproton momentum 
of 15~GeV in the design of HESR.
The central value of the cross section for $m_c = 1.5$~GeV and 
$\mu_r = \mu_f = 2 m_c$ at $\sqrt{s} = 5.47$~GeV is 12~nb.
The error band in Fig.~\ref{Fig:sig-mc} from varying the charm quark mass
corresponds to a multiplicative factor of about $7.4^2 \approx 55$.
The error band in Fig.~\ref{Fig:sig-mu} from varying $\mu_r$ and $\mu_f$  
corresponds to a multiplicative factor of about $3.1^2 \approx 9.6$.
The error band from varying $m_c$ is significantly wider.

\begin{figure}[t]
\includegraphics[width=0.8\textwidth,clip=true]{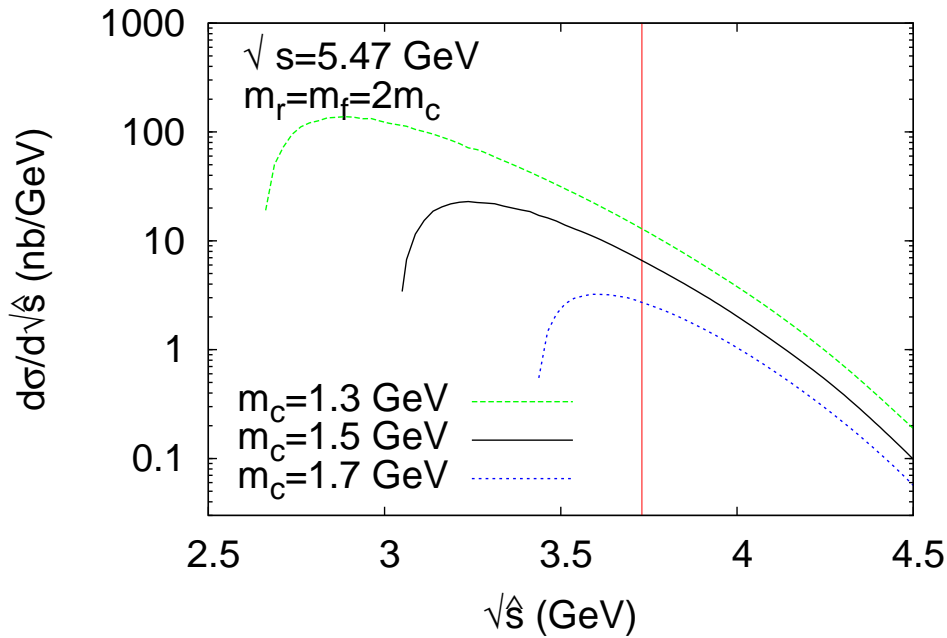}
\caption{Distribution of the invariant mass $\sqrt{\hat s}$ 
of the colliding partons at NLO for $\sqrt{s} = 5.47$~GeV,
$\mu_r = \mu_f = 2m_c$, and $m_c = 1.3$, 1.5, and 1.7~GeV 
(left, middle, and right curves).
The solid vertical line marks the position of the $D \bar D$ threshold.}
\label{Fig:s-hat}
\end{figure}

The reason for the large sensitivity to the charm quark mass is illustrated
in Fig.~\ref{Fig:s-hat}, which shows the distributions at NLO 
of the invariant mass $\sqrt{\hat s}$ of the colliding partons for 
$\sqrt{s} = 5.47$~GeV and $m_c = 1.3$, 1.5, and 1.7~GeV.
The full widths at half maximum of these distributions 
is about 500 MeV, which is large enough to provide the  
smearing required by our criterion {\bf A} 
for the validity of the parton model calculation.
However the peaks of the distributions are significantly below the 
$D \bar D$ threshold, which is indicated by the solid vertical line
in Fig.~\ref{Fig:s-hat}.  Thus our criterion {\bf B} is not satisfied.
This casts some doubt on the validity of the calculation for 
$\sqrt{s}$ as low as 5.47~GeV.

\begin{figure}[t]
\includegraphics[width=0.95\textwidth,clip=true]{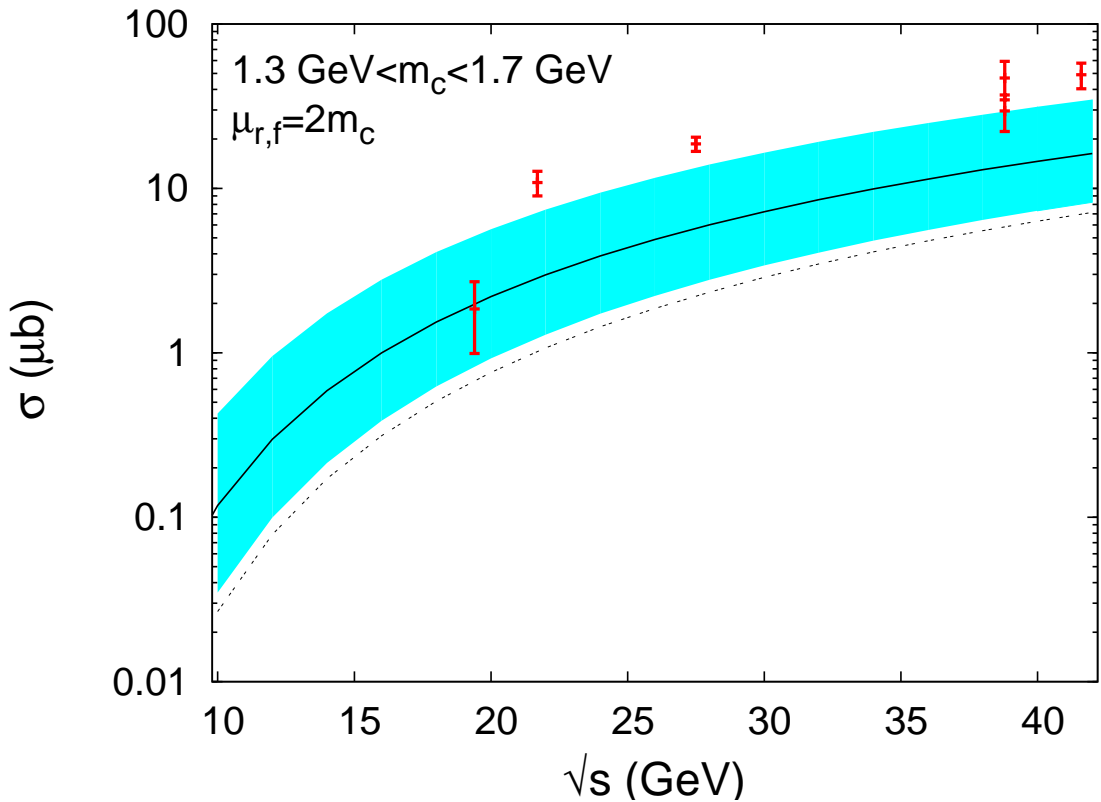}
\caption{Cross section for inclusive charm production in 
$p p$ collisions as a function of the center-of-mass energy
$\sqrt{s}$. The solid line is the NLO result for $m_c = 1.5$~GeV 
and $\mu_r = \mu_f = 2 m_c$.
The shaded band corresponds to varying $m_c$ from 1.3 and 1.7~GeV.
The dashed line is the LO result.  The data points are the measurements from 
Refs.~\cite{AguilarBenitez:1988sb,Barlag:1988qj,Ammar:1988ta,Kodama:1991jk,Alves:1996rz,Collaboration:2007zg}.} 
\label{Fig:sig-mc-pp}
\end{figure}

We can still regard the parton model calculation as a phenomenological model 
for the extrapolation of the charm cross section to low energies.
To reduce the model to one with a single parameter, 
we fix the scales at $\mu_r = \mu_f = 2 m_c$ and use the charm quark mass
$m_c$ as a phenomenological parameter.  
One way to determine $m_c$ is to require that parton model calculations give
the measured charm cross sections in low-energy fixed target experiments.
The most recent data on inclusive charm production from a proton beam
on a nuclear target with $p p$ center-of mass-energy below 45~GeV are
from the LEBC-MPS, ACCMOR, E743, E653, E769 and HERA-B
Collaborations~\cite{AguilarBenitez:1988sb,Barlag:1988qj,
Ammar:1988ta,Kodama:1991jk,Alves:1996rz,Collaboration:2007zg}.\footnote{ 
In Ref.~\cite{Frixione:1997ma}, 
the various charm meson cross sections published in 
Refs.~\cite{AguilarBenitez:1988sb,Barlag:1988qj,
Ammar:1988ta,Kodama:1991jk,Alves:1996rz} are converted into a common
cross section which is denoted by $\sigma_{c \bar c}$ but is actually
the inclusive cross section for $D^0$ and $D^+$ or, equivalently, 
the inclusive cross section for $\bar D^0$ and $D^-$.  To obtain the 
inclusive charm cross section, this must be divided by the sum of the
fragmentation probabilities for $c \to D^0$ and $c \to D^+$, which is
approximately 0.81.}
Their beam energies range from 200 up to 920~GeV, which corresponds to 
$pp$ center-of-mass energies from 19.4 up to 41.6~GeV.
The six data points are plotted in Fig.~\ref{Fig:sig-mc-pp} 
as a function of the center-of-mass energy $\sqrt{s}$.
Also shown in Fig.~\ref{Fig:sig-mc-pp} is the NLO calculation of the
inclusive charm cross section as a function of $\sqrt{s}$.  
The solid line is for $m_c = 1.5$~GeV, 
while the shaded band is the range as $m_c$ varies from 1.3 to 1.7~GeV.
For $\sqrt{s} = 19.4$~GeV and $m_c = 1.5$~GeV, 
the distribution of the invariant mass $\sqrt{\hat s}$ 
of the colliding partons has a peak near 4.1~GeV and a 
full width at half maximum of greater than 2 GeV.
Thus our criteria {\bf A} and {\bf B} for the validity 
of the parton model calculation are easily satisfied at this energy.
The six data points in Fig.~\ref{Fig:sig-mc-pp} 
are more compatible with the smaller values of $m_c$.
The best fit to the six data points is $m_c = 1.29$~GeV.

We proceed to use our results to assess the prospects for 
the study of charm meson mixing at HESR.
The maximum antiproton momentum is expected to be 15~GeV,
which corresponds to a center-of-mass energy of 5.47~GeV.
The NLO parton model with $m_c = 1.29$~GeV gives a cross section 
of $89$~nb at this energy. 
The design luminosity of HESR
is $2 \times 10^{32} /({\rm cm}^2 \, {\rm s})$.
Using the NLO cross section, we find that $5.6 \times 10^8$ charm events 
could be produced in a year of dedicated running.
Using our estimate $P_{c \to D^{*+}} = 24\%$ for the
fragmentation probability, we obtain 
$2.7 \times 10^8$ $D^{*+}$ or $D^{*-}$ events.  
The tagging of the initial flavor of the charm meson as 
$D^0$ or $\bar D^0$ comes from the decay into
$D^0 \pi^+$ or $\bar D^0 \pi^-$, which has a branching fraction of 68\%.
The simplest decay mode with which to observe mixing is the doubly-Cabbibo
suppressed decay mode $D^0 \to K^+ \pi^-$,
which has a branching fraction $1.5 \times 10^{-4}$.
Thus our estimate of the cross section corresponds to  
$27,000$  $D^0 \to K^+ \pi^-$  events.
To put this into context, the evidence for charm meson mixing
at the Tevatron from the CDF Collaboration used 13,000 
$D^0 \to K^+ \pi^-$ events \cite{CDF:2007uc}.
We conclude that the charm cross section at HESR
could be large enough to study charm meson mixing.

There is also a proposal to use the antiproton source for 
Fermilab's Tevatron for low-energy $\bar p p$ collisions \cite{Kaplan:2008pa}.
The maximum antiproton momentum in the 
accumulator ring at Fermilab is 8~GeV, which corresponds to 
a center-of-mass energy of only 4.11~GeV.  
The curves in Figs.~\ref{Fig:sig-mc} and \ref{Fig:sig-mu}
have been extended down to the $D \bar D$ threshold at 3.73~GeV, 
but this is too low for the model to be plausible,
because the cross section must vanish at this threshold.
A center-of-mass energy of 4.11~GeV may also be beyond the 
domain of plausibility of the model,
because it is only a little above the $D^* \bar D^*$ threshold.
If we ignore this problem, the NLO parton model with $m_c = 1.29$~GeV
gives an inclusive charm cross section of $4.1$~nb.

In summary, we have used the parton model and NLO perturbative QCD
to calculate the cross section for inclusive 
charm production by a low-energy antiproton beam.
At $\sqrt{s} = 5.47$~GeV, which corresponds to a $\bar p$ beam
with momentum 15 GeV, the cross section can be increased 
or decreased by an order of magnitude by varying $m_c$, $\mu_r$, 
and $\mu_f$ within reasonable ranges.
Since the fundamental justification for the calculation is questionable 
at such low energies, we have proposed the parton model calculation 
as a phenomenological model for extrapolating 
the charm cross section.  Treating $m_c$ as a phenomenological 
parameter, we determined its value by fitting measured charm
cross sections from fixed-target experiments with low-energy
proton beams.  The resulting estimate of the charm cross section 
at $\sqrt{s} = 5.47$~GeV is $89$~nb.  This cross section is 
large enough to allow the study of charm meson mixing in experiments 
with a 15~GeV antiproton beam.

P.A.\ was supported by the Fonds National de la Recherche
Scientifique and by the Belgian Federal Office for Scientific, 
Technical and Cultural Affairs through the 
Interuniversity Attraction Pole No.~P6/11.
E.B.\ was supported in part by the Department of Energy
under grant DE-FG02-91-ER40690 
and by the Alexander von Humboldt Foundation.


\end{document}